# Security Analysis of Secure Force Algorithm for Wireless Sensor Networks

## performance evaluation of 64, 128 and 192-bit secure force algorithm architecture

Shujaat Khan
Iqra University, Karachi, Pakistan
shujaat@iqra.edu.pk

M. Sohail Ibrahim
Iqra University, Karachi, Pakistan
msohail@iqra.edu.pk

Kafeel Ahmed Khan
Iqra University, Karachi, Pakistan
kafeel.khan@rocketmail.com

Mansoor Ebrahim
Sunway University, Selangor, Malaysia
12032389@imail.sunway.edu.my

***Abstract*— In Wireless Sensor Networks, the sensor nodes are battery powered small devices designed for long battery life. These devices also lack in terms of processing capability and memory. In order to provide high confidentiality to these resource constrained network nodes, a suitable security algorithm is needed to be deployed that can establish a balance between security level and processing overhead. The objective of this research work is to perform a security analysis and performance evaluation of recently proposed Secure Force algorithm. This paper shows the comparison of Secure Force 64, 128, and 192 bit architecture on the basis of avalanche effect (key sensitivity), entropy change analysis, image histogram, and computational time. Moreover, based on the evaluation results, the paper also suggests the possible solutions for the weaknesses of the SF algorithm.**

## I. Introduction

Modern advancement in the field of communication and computer networks increases the challenges for network security, scalability and reliability [16], [17]. Like all other communication networks wireless sensor networks are also prone to security issues. A WSN may contain several sensor nodes and each node consists of a processor, a limited battery power, memory, and communication ability. To ensure security in WSN, an algorithm that can provide optimum security with the resource constraints of WSN nodes is required. Conventional cryptographic algorithm is not suitable for WSN because of its distinctive characteristics [3]. The key issue in designing the cryptographic algorithms for WSN is to deal with the trade-off among security, memory, power, and performance. To achieve the high security requirements, numerous efforts have been made on assessing cryptographic algorithms and proposing energy efficient ciphers [4], [5] for WSN [6].

In our previous research work we proposed a low-complexity symmetric key algorithm for WSN, denoted as Secure Force (SF) and compare it with several existing symmetric key algorithms based on architecture, flexibility, and security level [1]. This paper shows the effect of increasing the key size of SF on security and computational complexity; we also performed test for key sensitivity and image encryption.

The rest of the paper is organized as; the introduction of Secure Force algorithm is discussed in Section 2. In section 3 the performance evaluation criteria are discussed. All the simulation results based on evaluation criteria are presented and discussed in Section 4. Finally, the conclusion is drawn in Section 5.

## II. Secure Force Algorithm

The Secure Force algorithm is based on a Fiestel architecture where the process of encryption and decryption are nearly the same, which minimizes the code size to a great extent. The design of SF algorithm provides low-complexity architecture for implementation in WSN. To improve the energy efficiency, the encryption process consists of only five encryption rounds. It has been suggested in [19] that a lower number of encryption rounds will result in less power consumption. In order to improve the security, each encryption round encompasses six simple mathematical operations operating on only 4 bit data (designed to be compatible with 8-bit computing devices for WSNs). This is to create an adequate amount of confusion and diffusion of data to encounter different types of attacks. The key expansion process, which involves complex mathematical operations (multiplication, permutation, transposition and rotation) to generate keys for the encryption process, is implemented at the decoder. This shifted the computational burden to the decoder and indirectly, this will help to increase the lifespan of the sensor nodes. However, the generated keys must be transmitted securely to the encoder for the encryption process.

In this case, the LEAP (Localized Encryption and Authentication Protocol) [18] is adopted. It is an energy efficient, robust and secure key management protocol that is designed for the WSN. Overall, the process of SF algorithm consists of 4 major blocks. The detail description of each block of the Secure Force algorithm can be found in [1]. The overall key transmission is depicted in Figure 1.

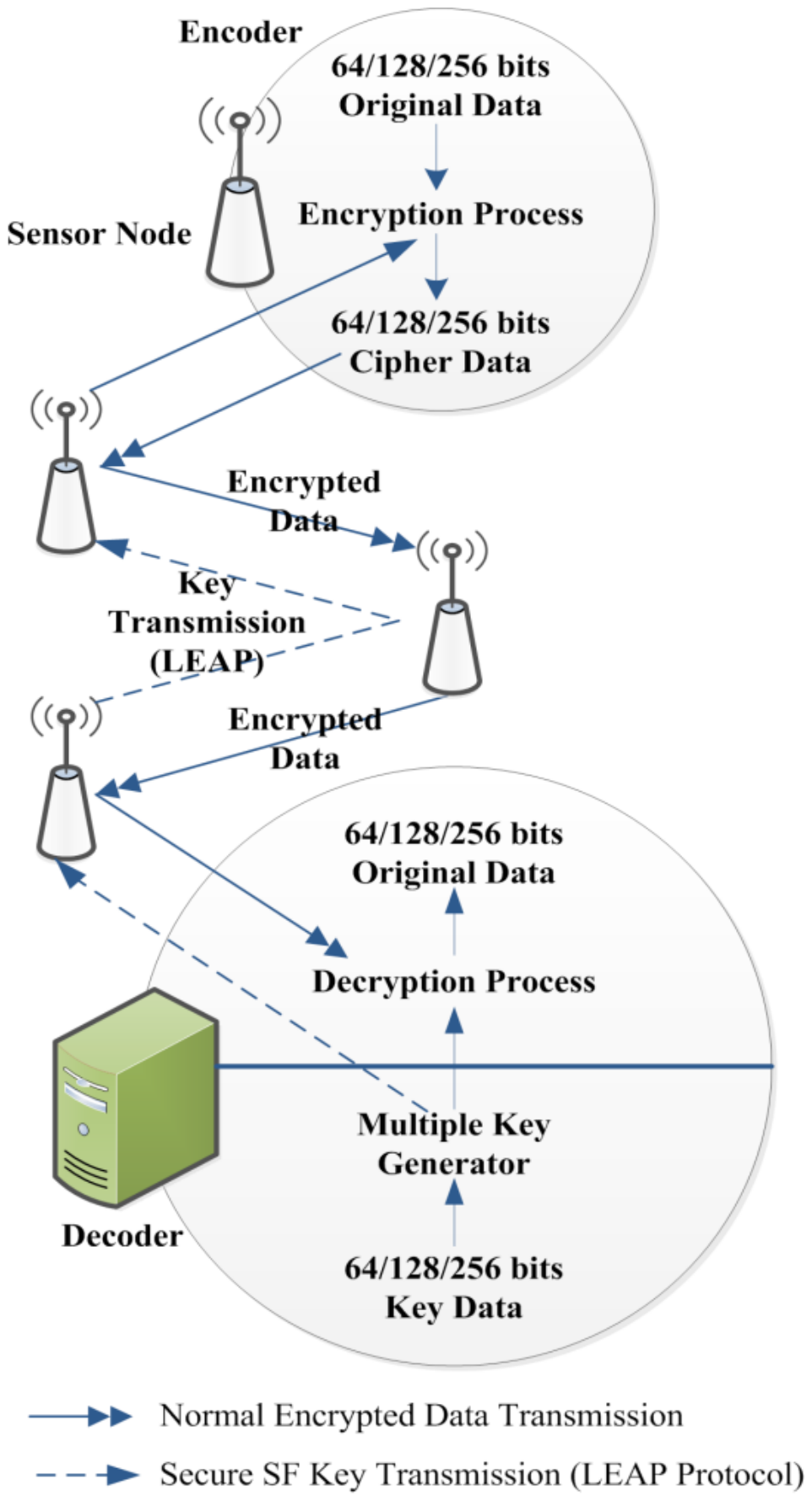

Figure 1. Key Transmission

- Key Expansion Block: Key expansion is the prime process that is used to generate different keys for encryption and decryption. Different operations are performed in order to create confusion and diffusion. This is to reduce the possibility of weak key as well as to increase the key strength. The round keys (Kr) are derived from the input cipher key by means of the key schedule. The process consists of two components: key expansion and round key selection. The key expansion performs logical operations (XOR, XNOR), left shifting (LS), matrix multiplication using fix matrix (FM), permutation using P-table and transposition using T-table.
- Key Management Protocol: The key can be securely sent to the encoder with the aid of LEAP [18].It is a simple and energy efficient protocol designed for large scale WSN, which allows secure key establishment through the use of four types of keys. They are known as the individual key, group key, cluster key, and pair wise shared key.
- Encryption Block: The encryption process is initiated once the keys generated by the key expansion block are securely received by the encoder through the secure communication channel created by the LEAP protocol. In the encryption process, simple operations, which include AND, OR, XOR, XNOR, left shift (LS), substitution (S boxes) and swapping operations, are performed to create confusion and diffusion.
- Decryption Block: The decryption process is just the reserve of the encryption process described above.

## III. Performance Evaluation Criteria

The evaluation of SF algorithm was carried out on certain well know parameters used by various authors [11], [12], [13], [14] and [15] in order to assess the performance of different conventional algorithms.

### A. *Avalanche Test*

A very well-known parameter used to analyze the security (randomness) of an encryption algorithm. The avalanche test measures the effect of change in the number of bits of encrypted text (cipher) due to one bit change in either key or plain text. The avalanche test is considered best if half of the bits of the cipher text are changed as per the strict avalanche criterion SAC [2].

### B. *Execution Time (Encoding/Decoding)*

The execution time is one of the essential parameter that needs to be considered along with security in the development of an encryption algorithm. The execution time of an encryption algorithm is defined as the total time required for the encoding/decoding of a particular data.

### C. *Image histogram*

Image histogram is a recently used parameter; it shows the randomness in the encrypted image data distribution. In this parameter the histogram of encrypted and unencrypted images are compared to know overall change in the data image intensities due to encryption.

### D. *Image Entropy*

Digital Images are combination of discrete valued pixels, combined together to form a visual perception of image. Image entropy measure is the simplest parameter used to analyze the randomness in the encrypted image. In this parameter the difference between the original and encrypted image’s entropy is measured. The greater the entropy change, the better will be the encryption. Entropy of an image can be calculated by the given relation (1)

$$E = \sum_{i=1}^{N} X_i(\log2(X_i)) \qquad (1)$$

Where ‘E’ is entropy of image, ‘X’ is the probability of the intensity level in image and ‘N’ is the total number of intensity levels.

## IV. Experimental Results

The experiments are performed on plain text as well as on image data. The original and encrypted images for SF-64, SF-128, and SF-192 are shown in Figure 2. The detailed description of the experiments and their results are discussed below.

### A. *Avalanche Test*

The results in the table1 show that Secure Force algorithm can cause a significant number of bits change with the single bit change in the key or plain text. SF 64, 128 and 192 can change 58.2%, 51.55%, and 45.70% respectively of cipher bits due to the change of one bit in text or key bits where as the avalanche

results of DES-64 is 65.63%[23] and for AES-128 it is 44.92%[12].

TABLE 1A. AVALANCHE TEST RESULTS FOR SF-64

| SNO. | KEY | INPUT | OUTPUT | AVALANCHE TEST |
|---|---|---|---|---|
| 1 | FFFFFFFFF<br>FFFFFFF | 000A4A6DE8<br>DB6667 | 925BEDEAD<br>4E631EB | 0.6250 |
| | FFFFF7FFF<br>FFFFFFF | 000A4A6DE8<br>DB6667 | B83D3E9D07<br>911E50 | |
| 2 | FFFFFFFFF<br>FFFFFFF | 000A4A6DE8<br>DB6667 | 925BEDEAD<br>4E631EB | 0.5156 |
| | FFFFFF7FF<br>FFFFFFF | 000A4A6DE8<br>DB6667 | 4D2CF304C7<br>D5E1EB | |
| 3 | 000A4A6DE<br>8DB6667 | FFFFFFFFFF<br>FFFFFF | 388D9772977<br>2388D | 0.5938 |
| | 000A4A6DE<br>8DB6667 | FFFFFFFFFF<br>FFDFFF | D5501BBEE5<br>D8F550 | |
| 4 | 000A4A6DE<br>8DB6667 | FFFFFFFFFF<br>FFFFFF | 388D9772977<br>2388D | 0.5938 |
| | 000A4A6DE<br>8DB6667 | FFFFFFFFEF<br>FFFFFF | E772F99D6D<br>D8288D | |
| Mean percentage avalanche value | | | | **0.5820** |

TABLE 2B. AVALANCHE TEST RESULTS FOR SF-128

| SNO. | KEY | INPUT | OUTPUT | AVALANCHE TEST |
|---|---|---|---|---|
| 1 | FFFFFFFFF<br>FFFFFFFFF<br>FFFFFFFFF<br>FFFFF | 000A4A6DE8<br>DB6667000A<br>4A6DE8DB66<br>67 | 3F8BB2B125<br>76CDD8D0B<br>CFFDA480C<br>EFDA | 0.5546 |
| | FFFFFFFFF<br>FFFFBFFFF<br>FFFFFFFFF<br>FFFFF | 000A4A6DE8<br>DB6667000A<br>4A6DE8DB66<br>67 | F0F2FAF95A<br>C19F9A69C2<br>15B22CF3260<br>2 | |
| 2 | FFFFFFFFF<br>FFFFFFFFF<br>FFFFFFFFF<br>FFFFF | 000A4A6DE8<br>DB6667000A<br>4A6DE8DB66<br>67 | 3F8BB2B125<br>76CDD8D0B<br>CFFDA480C<br>EFDA | 0.5546 |
| | FFFFFFFFF<br>FEFFFFFFF<br>FFFFFFFFF<br>FFFFF | 000A4A6DE8<br>DB6667000A<br>4A6DE8DB66<br>67 | F0F2FAF95A<br>C19F9A69C2<br>15B22CF3260<br>2 | |
| 3 | 000A4A6DE<br>8DB666700<br>0A4A6DE8<br>DB6667 | FFFFFFFFFF<br>FFFFFFFFFF<br>FFFFFFFFFF<br>FF | 5F371C254B6<br>8053C4B6805<br>3C5F371C25 | 0.4765 |
| | 000A4A6DE<br>8DB666700<br>0A4A6DE8<br>DB6667 | FFFFFFFFFF<br>FFFFFFF7FF<br>FFFFFFFFFF<br>FF | 244C654C95F<br>4DE670871D<br>1EC8B331825 | |
| 4 | 000A4A6DE<br>8DB666700<br>0A4A6DE8<br>DB6667 | FFFFFFFFFF<br>FFFFFFFFFF<br>FFFFFFFFFF<br>FF | 5F371C254B6<br>8053C4B6805<br>3C5F371C25 | 0.4765 |
| | 000A4A6DE<br>8DB666700<br>0A4A6DE8<br>DB6667 | FFFFFFFFF7<br>FFFFFFFFFF<br>FFFFFFFFFF<br>FF | 8B331825087<br>1D1EC95F4D<br>E67244C654C | |
| Mean percentage avalanche value | | | | **0.5155** |

TABLE 3C. AVALANCHE TEST RESULTS FOR SF-192

| SNO. | KEY | INPUT | OUTPUT | AVALANCHE TEST |
|---|---|---|---|---|
| 1 | FFFFFFFFF<br>FFFFFFFFF<br>FFFFFFFFF<br>FFFFFFFFF<br>FFFFFFFFF<br>FFF | 000A4A6DE8<br>DB6667000A<br>4A6DE8DB66<br>67000A4A6D<br>E8DB6667 | 790D56B1343<br>AF789D8096<br>E269D46CB4<br>CA24141F05<br>A2E4810 | 0.4531 |
| | FFFFFFFBF<br>FFFFFFFFF<br>FFFFFFFFF<br>FFFFFFFFF<br>FFFFFFFFF<br>FFF | 000A4A6DE8<br>DB6667000A<br>4A6DE8DB66<br>67000A4A6D<br>E8DB6667 | CC18D8B8C4<br>A40D1E2369<br>48A4F6B5806<br>8D4453CC84<br>2BEC110 | |
| 2 | FFFFFFFFF<br>FFFFFFFFF<br>FFFFFFFFF<br>FFFFFFFFF<br>FFFFFFFFF<br>FFF | 000A4A6DE8<br>DB6667000A<br>4A6DE8DB66<br>67000A4A6D<br>E8DB6667 | 790D56B1343<br>AF789D8096<br>E269D46CB4<br>CA24141F05<br>A2E4810 | 0.4375 |
| | FFFFFFFFF<br>FFDFFFFFF<br>FFFFFFFFF<br>FFFFFFFFF<br>FFFFFFFFF<br>FFF | 000A4A6DE8<br>DB6667000A<br>4A6DE8DB66<br>67000A4A6D<br>E8DB6667 | DEF94CD142<br>3C444DEE51<br>08A21BF1376<br>3F095A7C7F<br>A3E4910 | |
| 3 | 000A4A6DE<br>8DB666700<br>0A4A6DE8<br>DB6667000<br>A4A6DE8D<br>B6667 | FFFFFFFFFF<br>FFFFFFFFFF<br>FFFFFFFFFF<br>FFFFFFFFFF<br>FFFFFFFF | 7CE2BDAFD<br>041B2551593<br>5381B255159<br>353817CE2B<br>DAFD041 | 0.4688 |
| | 000A4A6DE<br>8DB666700<br>0A4A6DE8<br>DB6667000<br>A4A6DE8D<br>B6667 | FFFFFFFFFF<br>FFFFFFFFFF<br>FFFFFFFFFF<br>FFFFFFEFFF<br>FFFFFFFF | 2026C2A8501<br>1388BE628D<br>819E0A83621<br>58819FAD24<br>06DAD1 | |
| 4 | 000A4A6DE<br>8DB666700<br>0A4A6DE8<br>DB6667000<br>A4A6DE8D<br>B6667 | FFFFFFFFFF<br>FFFFFFFFFF<br>FFFFFFFFFF<br>FFFFFFFFFF<br>FFFFFFFF | 7CE2BDAFD<br>041B2551593<br>5381B255159<br>353817CE2B<br>DAFD041 | 0.4688 |
| | 000A4A6DE<br>8DB666700<br>0A4A6DE8<br>DB6667000<br>A4A6DE8D<br>B6667 | EFFFFFFFFF<br>FFFFFFFFFF<br>FFFFFFFFFF<br>FFFFFFFFFF<br>FFFFFFFF | 9FAD2406DA<br>D1E0A83621<br>5881388BE62<br>8D8192026C2<br>A85011 | |
| Mean percentage avalanche value | | | | **0.4570** |

*B. Simulation Time*

SF is a light weight algorithm, it uses very low amount of computer resources. The results in table 2 show overall computation time for encryption. The comparison of SF with AES algorithm on FPGA platform is mentioned in [24]. When implemented on MATLAB®, SF-192 with 180mSec is the most expensive choice in terms of computational cost among the three versions. SF-64 and SF-128 take 27.5mSec and 35mSec simulation time respectively. Although SF shows good results and its performance is comparable to other algorithms [7], [8], [9], [10] and [15] in terms of computation cost, but it is still not as claimed.

TABLE 4. SIMULATION TIME ANALYSIS AND EXPERIMENTAL SETUP

| Architecture | **SF-64** | **SF-128** | **SF-192** |
|---|---|---|---|
| Mean Simulation Time | 27.5mSec | 35mSec | 180mSec |
| Processor | Intel Core2Duo T6500 @ 2.10GHz | | |
| RAM | 4.00 GB | | |
| Operating System | Windows 8 64-bit | | |
| MATLAB Version | 7.12.0 R2011a 64-bit | | |

### *C. Image Histogram (Intensity Variation)*

Histogram is a very useful way to analyze the effect of encryption over the image. The ideal resultant histogram after encryption should be a straight line. SF shows pretty decent results for 128-bit architecture. For few test images, minor changes are observed in the histogram; this is due to their original intensity distribution. The results for selected images are shown in Figure 3.

### *D. Image Entropy*

Entropy is the measure of information contents of the data, more random the data more difficult it is to be recognized after encryption. The entropy change for six popular images, namely Cameraman [22], Rice [22], Lena [21], Football [22], ORL Faces [20], and Onion [22] is presented in table 3. On average 10.39%, 11.44%, and 11.33% of entropy change is observed with SF-64, SF-128, and SF-192 respectively. From the results, it is evident that SF-128 is the most appropriate choice for image encryption among the three versions of SF.

TABLE 5A. IMAGE ENTROPY TEST FOR SF-64

| **SNO.** | **Image** | **Dimension** | **Entropy (Original)** | **Entropy (Encrypted)** | **Percent Change** |
|---|---|---|---|---|---|
| 1 | Cameraman.tif | 256X256 | 7.0097 | 7.8705 | 12.28 |
| 2 | Rice.tif | 256X256 | 7.0115 | 7.9448 | 13.31 |
| 3 | Lena.jpg | 220X220 | 7.4618 | 7.9643 | 6.73 |
| 4 | Football.jpg | 256X320 | 6.6861 | 7.8210 | 16.97 |
| 5 | ORLFace.jpg | 490X467 | 7.5332 | 7.9723 | 5.83 |
| 6 | Onion.png | 135X198 | 7.3299 | 7.9300 | 8.19 |
| Mean Entropy values | | | **7.1720** | **7.9172** | **10.39** |

TABLE 6B. IMAGE ENTROPY TEST FOR SF-128

| **SNO.** | **Image** | **Dimension** | **Entropy (Original)** | **Entropy (Encrypted)** | **Percent Change** |
|---|---|---|---|---|---|
| 1 | Cameraman.tif | 256X256 | 7.0097 | 7.9927 | 14.02 |
| 2 | Rice.tif | 256X256 | 7.0115 | 7.9923 | 13.99 |
| 3 | Lena.jpg | 220X220 | 7.4618 | 7.9945 | 7.14 |
| 4 | Football.jpg | 256X320 | 6.6861 | 7.9917 | 19.53 |
| 5 | ORLFace.jpg | 490X467 | 7.5332 | 7.9973 | 6.16 |
| 6 | Onion.png | 135X198 | 7.3299 | 7.9847 | 8.93 |
| Mean Entropy values | | | **7.1720** | **7.9922** | **11.44** |

TABLE 7C. IMAGE ENTROPY TEST FOR SF-192

| **SNO.** | **Image** | **Dimension** | **Entropy (Original)** | **Entropy (Encrypted)** | **Percent Change** |
|---|---|---|---|---|---|
| 1 | Cameraman.tif | 256X256 | 7.0097 | 7.9878 | 13.95 |
| 2 | Rice.tif | 256X256 | 7.0115 | 7.9834 | 13.86 |
| 3 | Lena.jpg | 220X220 | 7.4618 | 7.9940 | 7.13 |
| 4 | Football.jpg | 256X320 | 6.6861 | 7.9658 | 19.14 |
| 5 | ORLFace.jpg | 490X467 | 7.5332 | 7.9971 | 6.15 |
| 6 | Onion.png | 135X198 | 7.3299 | 7.9774 | 8.83 |
| Mean Entropy values | | | **7.1720** | **7.9842** | **11.33** |

## V. CONCLUSION

Power constrained network like wireless sensor networks (WSN) demands an algorithm which can provide a reliable security at an affordable computational cost. Recently proposed Secure Force algorithm is one of the candidates for WSN security solutions. In this paper we implemented the SF (64, 128 and 192 bit) architectures on MATLAB® platform and perform the various standard tests for image and text data. The test results show that SF performs reasonably well in terms of computational time and randomness. In the results of security analysis of SF (64-bit) many weaknesses were identified which are rectified in the 128 and 192-bit architecture. After the extensive testing under the strict performance evaluation criteria, it is concluded that SF-128 performs exceptionally well when compared with SF-64 and SF-192. SF-128 stands first in the list with 51.55% Avalanche value, 35mSec simulation time for Encryption/Decryption, and 11.44% Entropy change.

## Image encryption results

**Orignal** | **Encrypted with SF(64)** | **Encrypted with SF(128)** | **Encrypted with SF(192)**

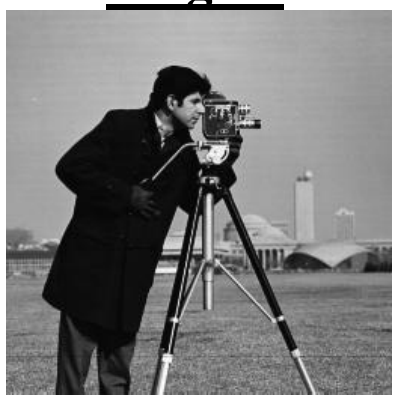

Fig. 2.1a : Cameraman original image

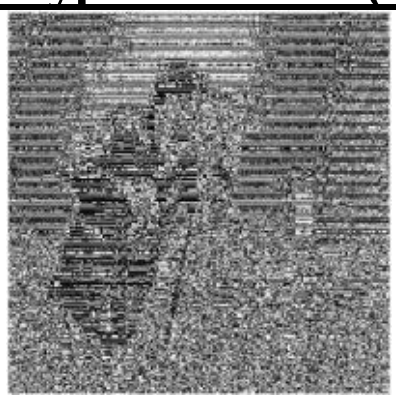

Fig. 2.1b : Cameraman encrypted image

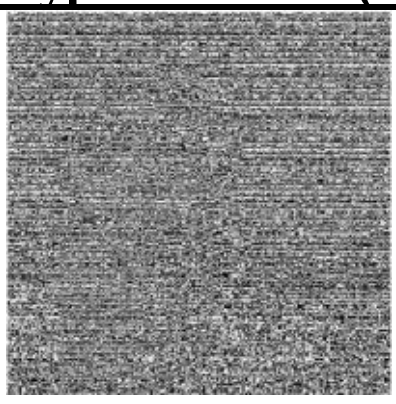

Fig. 2.1c : Cameraman encrypted image

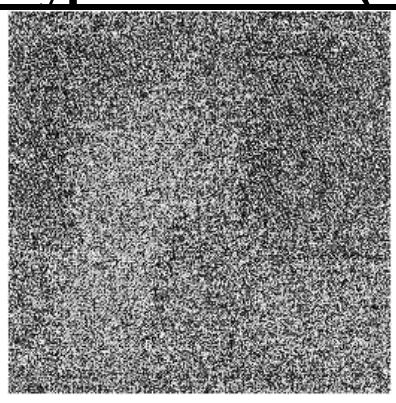

Fig. 2.1d : Cameraman encrypted image

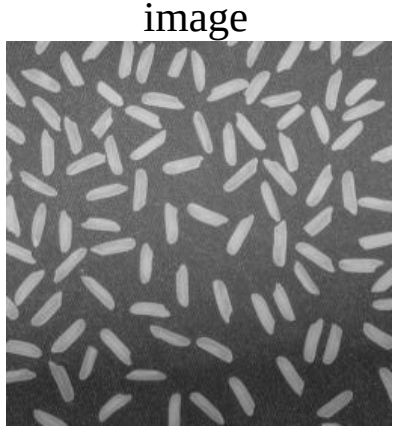

Fig. 2.2a : Rice original image

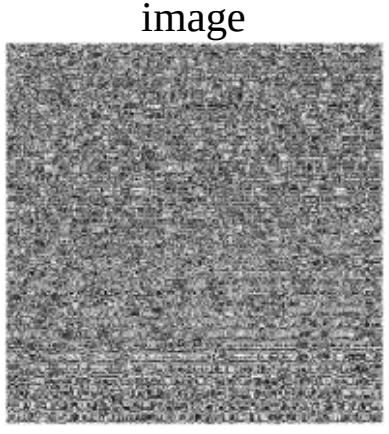

Fig. 2.2b : Rice encrypted image

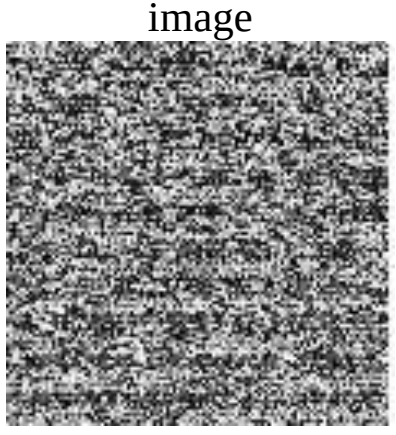

Fig. 2.2c : Rice encrypted image

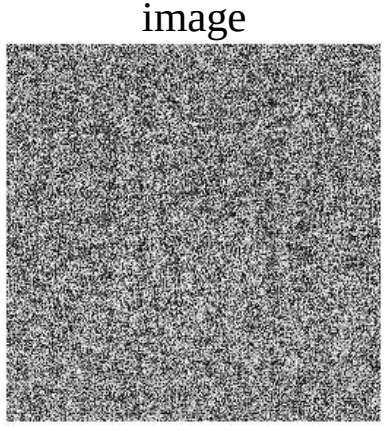

Fig. 2.2b : Rice encrypted image

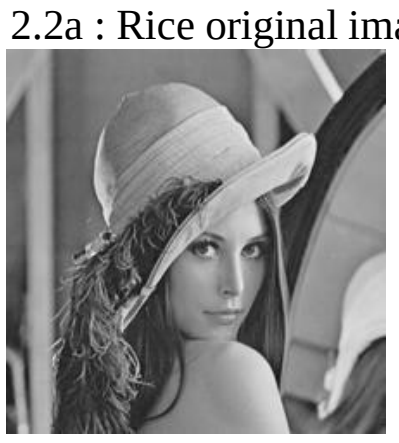

Fig. 2.3a : Lena original image

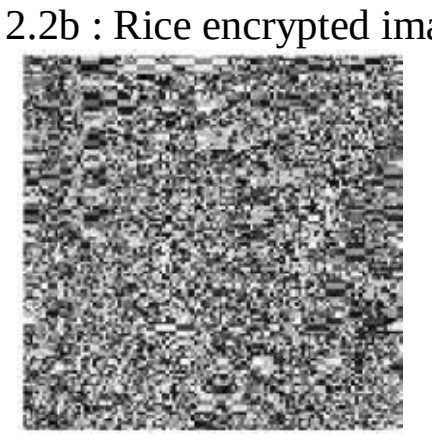

Fig. 2.3b : Lena encrypted image

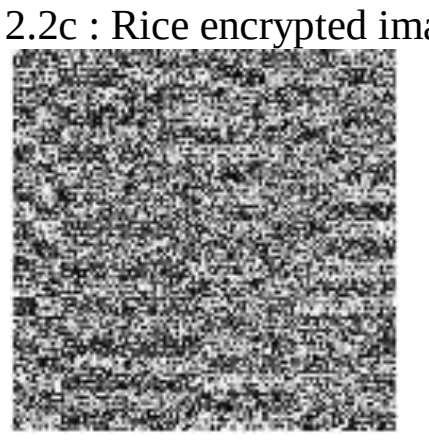

Fig. 2.3c : Lena encrypted image

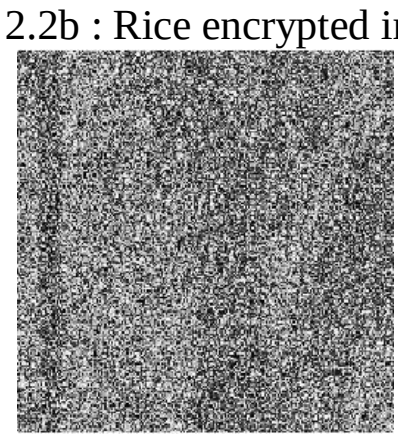

Fig. 2.3d : Lena encrypted image

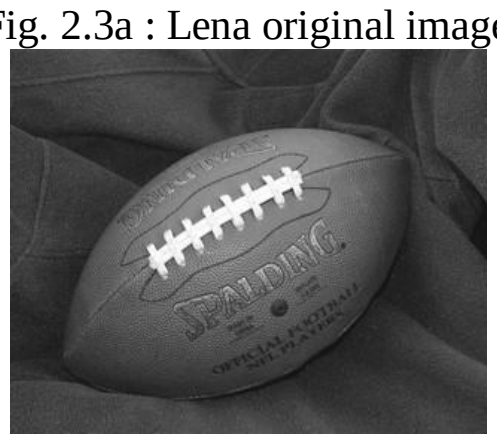

Fig. 2.4a : Football original image

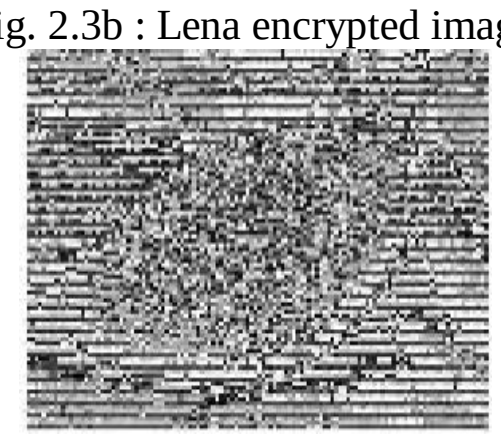

Fig. 2.4b : Football encrypted image

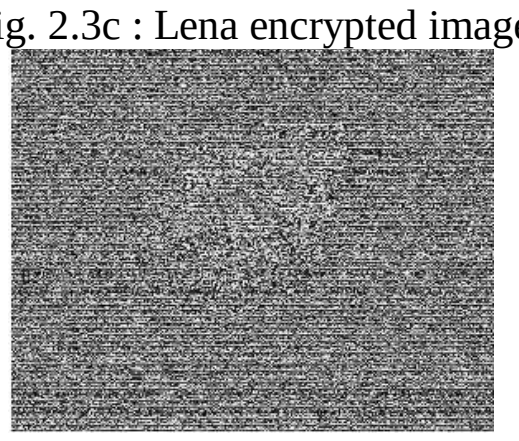

Fig. 2.4c : Football encrypted image

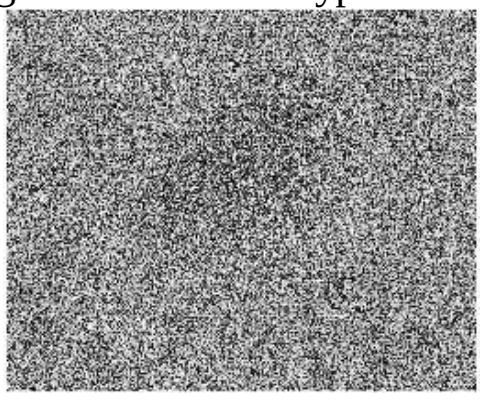

Fig. 2.4d : Football encrypted image

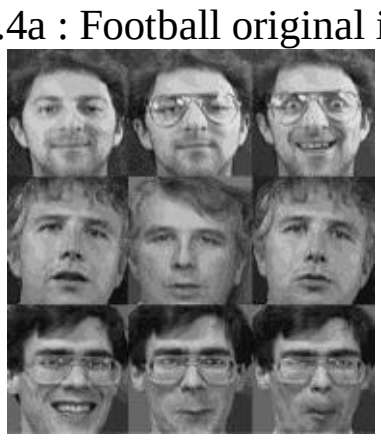

Fig. 2.5a : ORL Faces original image

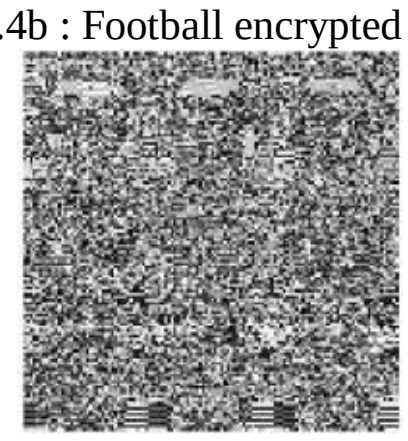

Fig. 2.5b : ORL Faces encrypted image

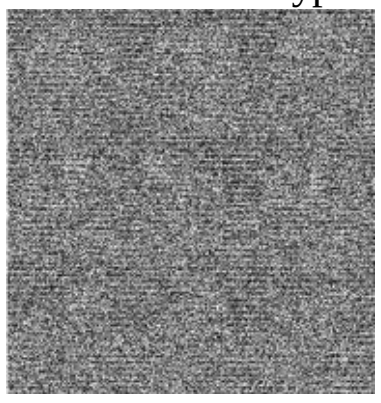

Fig. 2.5c : ORL Faces encrypted image

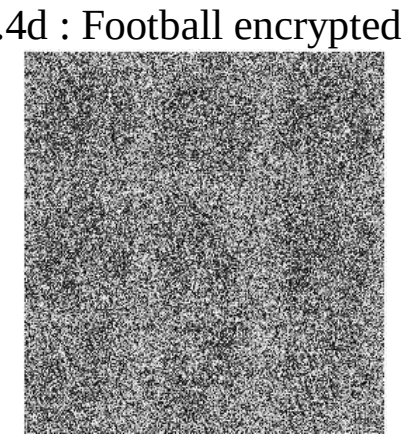

Fig. 2.5d : ORL Faces encrypted image

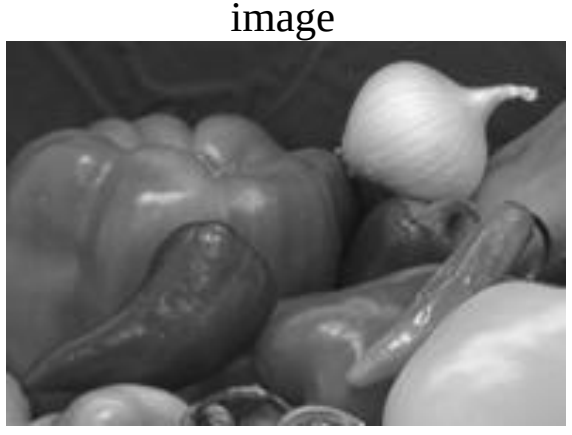

Fig. 2.6a : Onion original image

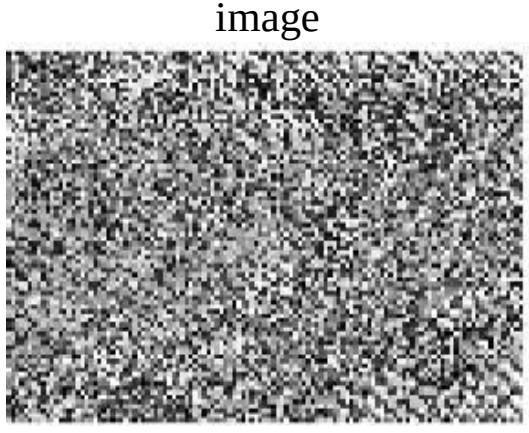

Fig. 2.6b : Onion encryption image

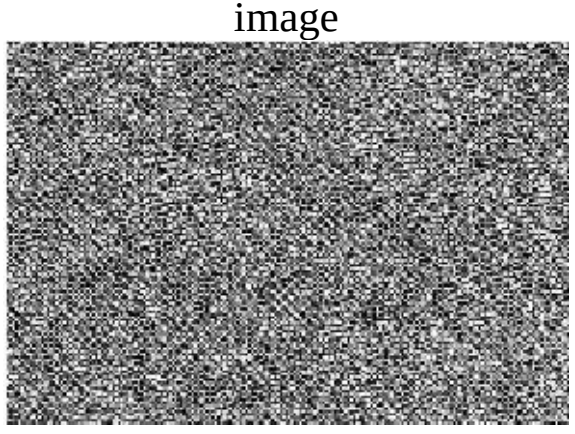

Fig. 2.6c : Onion encryption image

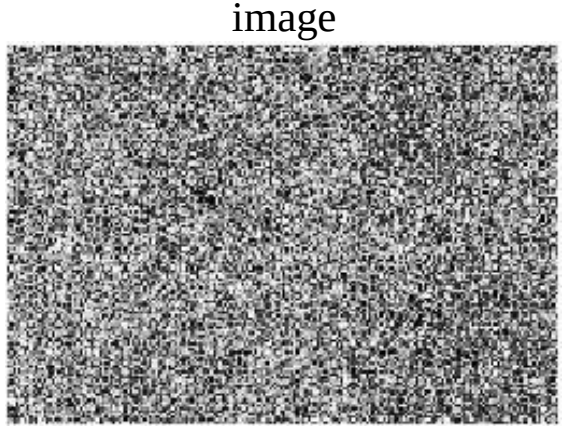

Fig. 2.6d : Onion encryption image

## Image histogram results

### Image histogram results for SF-64

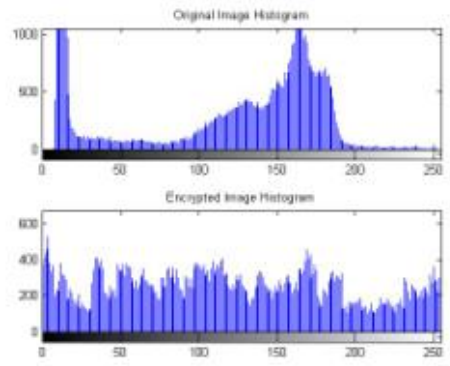

Fig.3.1a : Histogram of "Cameraman"

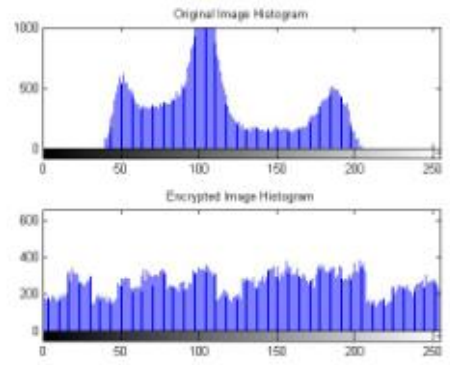

Fig. 3.2a : Histogram of “rice”

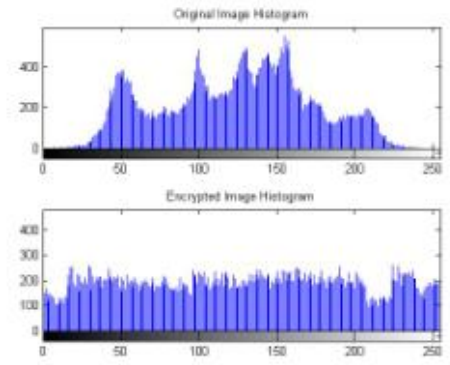

Fig. 3.3a : Histogram of “Lena”

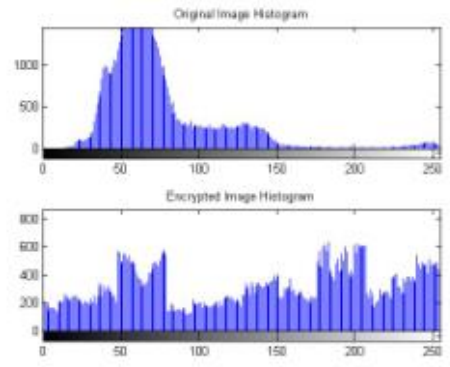

Fig. 3.4a : Histogram of “Football”

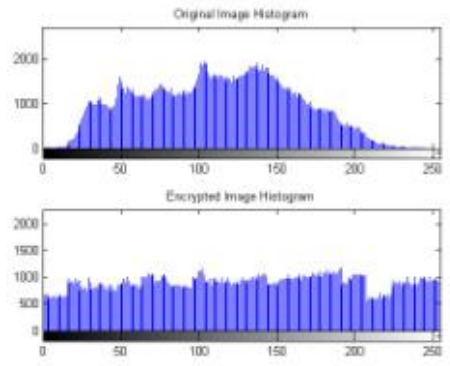

Fig. 3.5a : Histogram of “ORL Faces”

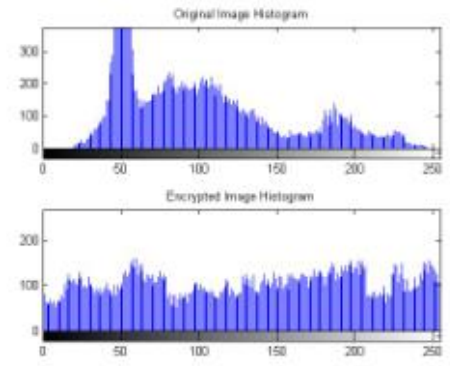

Fig. 3.6a : Histogram of “Onion”

### Image histogram results for SF-128

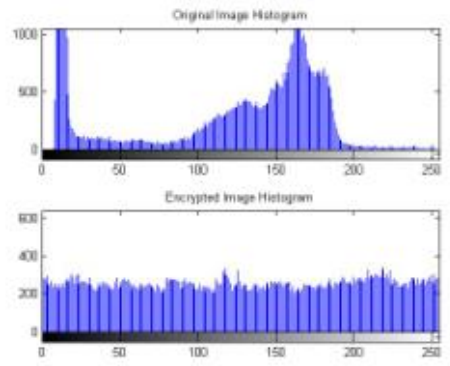

Fig. 3.1b : Histogram of "Cameraman"

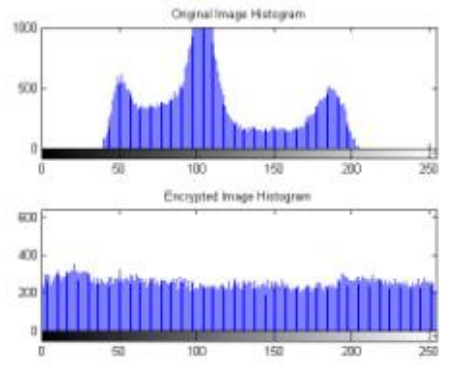

Fig. 3.2b : Histogram of “rice”

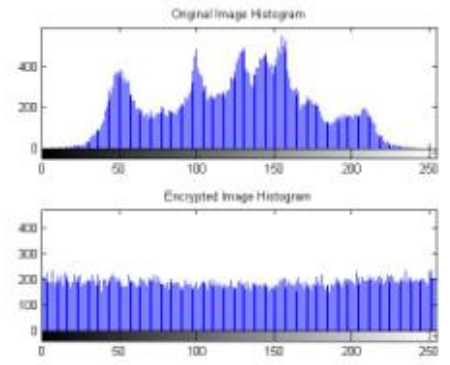

Fig. 3.3b : Histogram of “Lena”

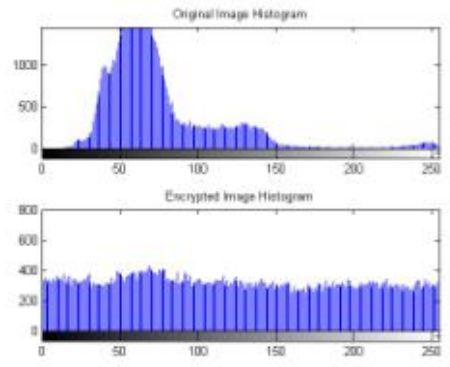

Fig. 3.4b : Histogram of “Football”

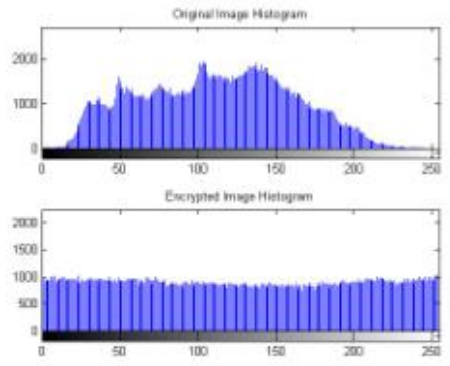

Fig. 3.5b : Histogram of “ORL Faces”

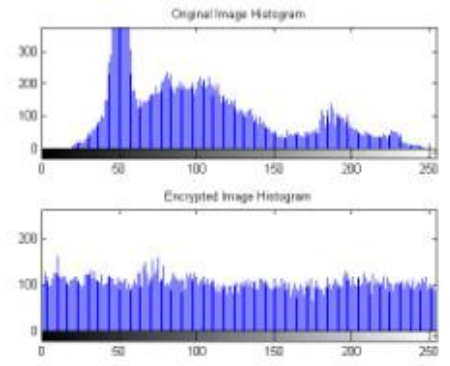

Fig. 3.6b : Histogram of “Onion”

### Image histogram results for SF-192

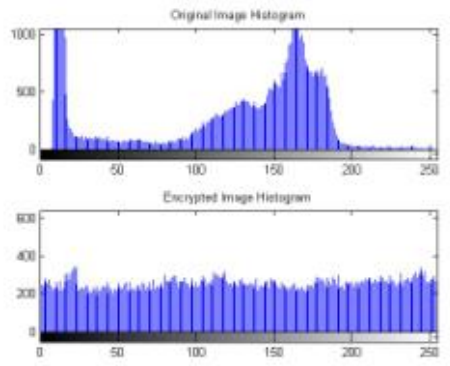

Fig.3.1c : Histogram of "Cameraman"

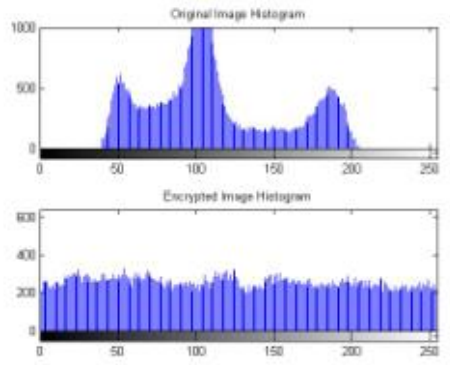

Fig. 3.2c : Histogram of “rice”

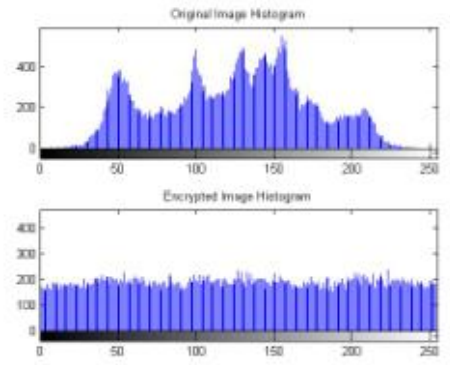

Fig. 3.3c : Histogram of “Lena”

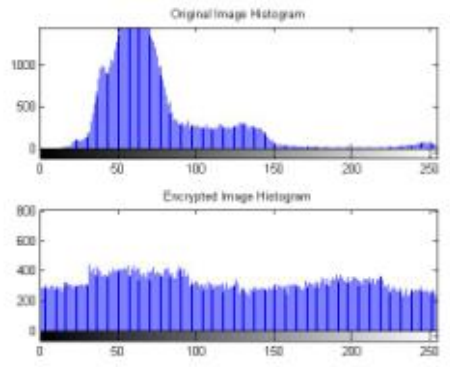

Fig. 3.4c : Histogram of “Football”

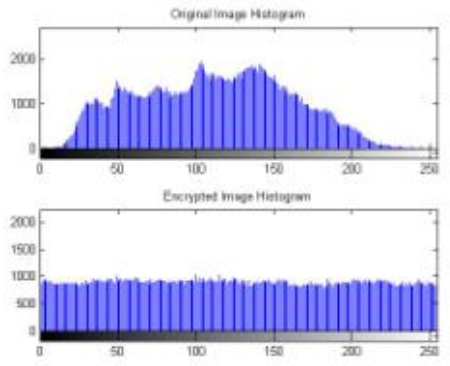

Fig. 3.5c : Histogram of “ORL Faces”

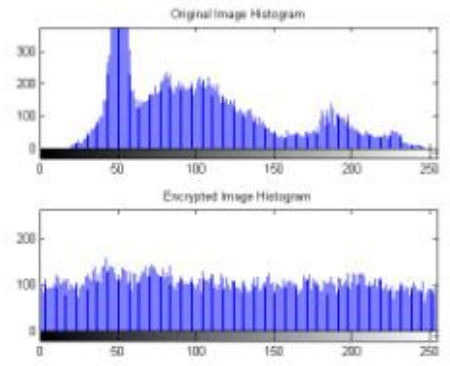

Fig. 3.6c : Histogram of “Onion”